\documentstyle[12pt]{article}
\renewcommand{\theequation}{\arabic{section}.\arabic{equation}}

\begin{document}
\def\l{\lambda}

\begin{flushright}
UT-Komaba 97-14   \\
\end{flushright}

\begin{center}
{\Large{\bf  Low-Energy Excitations in 2-Leg and 3-Leg}}
\vskip 0.3cm
{\Large{\bf  Quantum Spin Ladders }} 
\vskip 1cm
{\Large Ikuo Ichinose\footnote{e-mail address:
ikuo@hep1.c.u-tokyo.ac.jp} and Yasuyuki Kayama\footnote{e-mail address:
kayama@amon.c.u-tokyo.ac.jp}}

\vskip 0.3cm
Institute of Physics, University of Tokyo, Komaba, Tokyo, 153 Japan

\end{center}
\vskip 0.3cm
\begin{center} 
\begin{bf}
Abstract
\vskip 0.5cm
\end{bf}
\end{center} 
Low-energy excitations in spin-${1 \over 2}$ antiferromagnetic(AF) Heisenberg
spin ladders are studied by bosonization and gauge -theoretical description.
The AF Heisenberg models on ladders are described by spin-${1 \over 2}$
fermions and the systems are reduced to relativistic gauge theories of fermions
by a mean-field type decoupling and linearization of fermion's dispersion relation
near Fermi points.
There gauge field is nothing but phase degrees of freedom of ``mean field" on links.
It is explicitly shown that zero-mode part of boson fields for the bosonization of fermions 
plays an essentially important role to obtain correct results.
In the 2-leg case, the lowest-energy excitations are spin-triplet magnon and they are described by 
three massive Majorana fermions.
On the other hand in the 3-leg system, spin excitations on the top and bottom chains
are described by two massless boson fields.
It is predicted that if coupling between spins on the top and 
bottom chains is introduced,  a phase transition occures at some critical
coupling.
Above the critical coupling, the system acquires an energy gap and  low-energy excitations 
are spin-triplet magnon and spin-singlet excitation.

\newpage

\section{Introduction}
\setcounter{footnote}{0}

In the last few years, quasi-one-dimensional strongly correlated electron
systems are one of the most interesting topics in the condensed matter physics.
These systems are closely related to the high-T$_C$ cuprates and 
studies on them may shed light on mechanism of  the high-T$_C$ superconductivity.
Recently, however, it has been recognized that, besides relationship with
the high-T$_C$ cuprates, electron systems in quasi-one-dimensional space
exhibit very interesting behaviors themselves which are to be investigated\cite{rice}.  

In this paper, we shall study spin-${1 \over 2}$ antiferromagnetic(AF) Heisenberg models
on ladders.
At present it is well-established that quantum spin-${1 \over 2}$ systems
have gapless excitations on odd-number-leg ladders whereas only gapfull excitations 
on even-number-leg ladders.
This fact has been shown by numerical calculations and experiments,
and  understood by analytical
studies through the nonlinear-sigma model approach\cite{NLS}.

Bosonization is also useful for (quasi-)one-dimensional electron systems.
Among studies by the bosonization, Shelton, Nersesyan and Tsvelik showed
that massive excitation in the 2-leg spin ladder is described by massive
Majorana fermions\cite{shelton}.
They started with two decoupled Hubbard chains with belief that low-energy
excitations of the repulsive Hubbard chain  are described by the AF Heisenberg chain,
and then studied effects of the inter-chain interaction between spins by perturbation.
Later it was verified that their theory explaines experiments very well\cite{kishine}.
However, we encounter difficulties in deriving their results by straightforward
use of bosonization formulas.
Especially we cannot obtain their expressions for the staggered part of 
the local-spin density (Eq.(A18) in Ref.\cite{shelton}) which plays an 
essentially important role to derive the effective Hamiltonian of inter-ladder interaction
(Eqs.(7) and (8)).
This is one of our motivations to study the AF Heisenberg models on ladders
in more systematic and detailed manner.

We shall directly study the Heisenberg models though most of theoretical
studies are given for the Hubbard models\cite{hubbard}.
To this end we employ a gauge-theoretical description of the AF Heisenberg model
which is obtained by a mean-field type decoupling with keeping phase degrees of
freedom of ``mean fields" as dynamical variables.
Very recently similar investigation on 2-leg quantum spin ladder was given by
Hosotani\cite{hosotani}.
However, we think that results of that investigation are not physically acceptable.
This point will be discussed later on.

This paper is organized as follows.
In Sect.2, we shall give the gauge-theoretical description of the AF Heisenberg
model on 2-leg ladder.
Spin operators are expressed by spin-${1 \over 2}$ fermion bilinears and decoupling
of the intra-chain spin-spin interaction is performed.
We keep phase degrees of freedom of link auxiliary fields which are regarded as 
gauge fields.
We intergarte out high-momentum modes of fermions and verify that
Maxwell term of gauge fields appears.
Low-momentum effective theory is nothiing but a Schwinger model of multi-flavor fermions.
Then by using bosonization, we shall study effects of inter-chain
interaction rather in detail.
There zero modes of boson field for bosonization play an essential important role,
and because of them we derive the results in Ref.\cite{shelton}.

In Sect.3, we shall study the 3-leg quantum spin ladder.
Especially we forcus on low-energy excitations on the top and bottom
chains.
We use the same method of bosonization in the 2-leg case but we have to
introduce redundant boson fields in order to separate physical excitations.
To recover the original 3-leg spin system, we impose  local constraints.
We study how these constraints influence low-energy excitations by 
integrating out massive modes and show that two boson fields remain massless
and they describe low-energy excitations in the 3-leg spin ladder.
We also consider effects of spin-spin interactions (both ferromagnetic
and antiferromagnetic couplings) between spins on the top and bottom chains
and show that the system acquires an energy gap above some critical
coupling.
This prediction should be checked by numerical calculation etc.
 
Section 4 is devoted for discussion and conclusion.

 
\section{AF Heisenberg model on 2-leg ladder}
\setcounter{equation}{0}

Hamiltonian of the $l_c$-leg spin-ladder system is given as follows,
\begin{equation}
H_{l_c}=\sum_i\Big[J\sum_{r=1}^{l_c}\vec{S}_{i+1,r}\vec{S}_{i,r}
+J'\sum_{r=1}^{l_c-1}\vec{S}_{i,r}\vec{S}_{i,r+1}\Big],
\label{Hl}
\end{equation}
where site on $l_c$-leg ladder is labeled as $(i,r=1,...,l_c)$. 
The operator $\vec{S}_{i,r}$ is the spin operator with the magnitude ${1\over 2}$,
and we express it in terms of fermions 
$\psi_{i,r}=(\psi^{\alpha}_{i,r})$ $(\alpha=1,2 \;
\mbox{or} \uparrow, \downarrow)$,
\begin{equation}
\vec{S}_{i,r}={1 \over 2} \psi^{\dagger}_{i,r}\vec{\sigma}\psi_{i,r},
\label{spinop}
\end{equation}
with constraint
\begin{equation}
\sum_{\alpha}\psi^{\alpha\dagger}_{i,r}\psi^{\alpha}_{i,r}=1,
\label{constr}
\end{equation}
where $\vec{\sigma}$ is the Pauli spin matrices.
In the expression in terms of the fermions, the Hamiltonian is invariant under
a local ``gauge" transformation,
\begin{equation}
\psi_{i,r} \rightarrow e^{i\theta_{i,r}}\psi_{i,r}.
\label{gaugetrf1}
\end{equation}
Actually quantum spin systems can be described by gauge theories.  
This was first shown for the chain case in Ref.\cite{mukaida} and very recently
for ladder case in Ref.\cite{hosotani}.
However results in  Ref.\cite{hosotani} do not agree with those in Ref.\cite{shelton},
and then we shall revisit this system in this section.

The following identity is easily verfied 
$$
\vec{S}_i\cdot \vec{S}_j={1 \over 4}\Big[-\{\psi^{\dagger}_i\psi_j,\psi^{\dagger}_j\psi_i\}
+1-(\psi^{\dagger}_i\psi_i-1)(\psi^{\dagger}_j\psi_j-1)\Big].
$$
We consider the case $J/J' >1$ and start with decoupled Heisenberg chains.
For each chain we introduce auxiliary link field $U_{i,r}$ and Lagrangian of the
spin-chain system is given as (in the imaginary time formalism),
\begin{equation}
L^r_{chain}=\sum_i\Big\{ -\psi^{\dagger}_{i,r}\dot{\psi}_{i,r}-iA^{(r)}_{0}(i)(\psi^{\dagger}_{i,r}
\psi_{i,r}-1)-{J \over 2}\Big(U^{\dagger}_{i,r}U_{i,r}-U_{i,r}\psi^{\dagger}_{i,r}
\psi_{i+1,r}-U^{\dagger}_{i,r}\psi^{\dagger}_{i+1,r}\psi_{i,r}\Big)\Big\}.
\label{lag1}
\end{equation}
We can determine amplitude of the link field $U=|U_{i,r}|$ by the mean-field
calculation.
In this paper we consider chain and ladder of finite length $N$ and impose
periodic boundary condition for the spin variables 
$\vec{S}_{-{N \over 2},r}=\vec{S}_{{N \over 2},r}$.
By straightforward calculation we obtain $U={1 \over \pi}$ for sufficiently large $N$.
We shall take into account effect of phase degrees of freedom of  $U_{i,r}$ and 
parameterize it as
\begin {equation}
U_{i,r}={1 \over \pi} e^{iaA^{(r)}_1(i)},
\label{UA}
\end{equation}
where $a$ is the lattice spacing and $A^{(r)}_{\mu}=(A^{(r)}_0,A^{(r)}_1)$ is gauge field
for gauge transformation (\ref{gaugetrf1}).

For $U_{i,r}=U$ dispersion relation of the fermions is that of tight-binding model
and Fermi points are located at  $k=\pm k_F=\pm {\pi \over 2a}$ as the present
system is at half-filling by the constraint (\ref{constr}). 
We are going to focus on low-momentum excitations and integrate out high-momentum
modes of fermions.
To this end we introduce a momentum cut off parameter $\Lambda (\ll {\pi \over 2a})$ 
and employ the approximation of linear dispersion relation for the momentum region
$|k\pm k_F|<\Lambda$. 
For low-energy excitations near the Fermi points, we define right-moving
and left-moving fermions $(\psi_R \psi_L)$ as in the usual way.
By integrating over the high-momentum modes there appear renormalization of
Fermi velocity etc.
However the most important term for low-momentum excitations is the Maxwell term
of the gauge fields $A^{(r)}_{\mu}$.
Therefore we include this term.
After rescaling the fermion fields, the following 2-dimensional U(1) gauge theory
is obtained as a low-energy effective field theory of the quantum spin chain 
\begin{equation}
{\cal L}^{(r)}_{chain}=-{1 \over e^2}F^2_{\mu\nu}+\sum_{\alpha=1,2}\bar{\psi}^{\alpha}_r
\gamma_{\mu}D_{\mu}\psi^{\alpha}_r+{ 1 \over a} A^{(r)}_0, \;\; 
D_{\mu}=\partial_{\mu}+iA_{\mu},
\label{lag2}
\end{equation}
where the coupling constant $e^2$ is a function of $J$, $a$ and $\Lambda$,
\begin{eqnarray}
e^2&=&{4 \over 3}{J \over a}\Big[\int^{{\pi \over 2}-a\Lambda}_{0}
dx {1 \over \cos x}\Big]^{-1} ,  \nonumber   \\
&\sim &{4 \over 3}{J \over l}\Big({l \over a}{1 \over \ln ({\pi a \over l})}\Big), \nonumber  \\
{ \pi \over l} &\equiv& \Lambda,
\label{gcoupling}
\end{eqnarray}
and $\gamma$-matricies are explicitly given by the Pauli matrices as $\gamma^0=\sigma_1$, 
$\gamma^1=\sigma_2$ and therefore $\gamma^5=\sigma_3$.

The Schwinger model (\ref{lag2}) can be solved by bosonization technique\cite{boson}.
Bosonization formula is given as 
\begin{eqnarray}
  \psi_{r \nu \alpha}(t,x)
    &=& \frac{1}{\sqrt{L}}
      \, C_{r \nu \alpha} \,
      e^{\pm i \{ q_{r \nu \alpha} + 2\pi p_{r \nu \alpha} (t \pm x)/L \} }
       N_0[ e^{\pm i\sqrt{4\pi}\phi_{r \nu \alpha} (t,x) } ],     \nonumber \\
\phi_{r \nu \alpha} (t,x) &=& \sum_{n=1}^{\infty}\Big( a^{\alpha}_{r  \nu,n}e^{-2\pi in
(t \pm x)/L} +\mbox{h.c.}\Big),   \quad  (L=Na, \quad \nu=L,R )
\label{bosonizedF1}
\end{eqnarray}
where $N_m$ is the normal-ordering operator with mass $m$ and
commutation relations are 
$ [q_{r\nu\alpha}, p_{r'\nu'\alpha'}]=i \delta_{rr'}\delta_{\nu\nu'}\delta_{\alpha\alpha'},
\; [a^{\alpha}_{r\nu,n},a^{\alpha'}_{r'\nu',m}]=\delta_{rr'}
\delta_{\alpha\alpha'}\delta_{\nu\nu'}\delta_{nm}$ 
and $C$'s are functions of $p$'s
which make the fermion fields $\psi_{r \nu \alpha}(t,x)$ satisfy the canonical
anticommutation relations (CAR).

In the rest of this section, we shall consider the two-leg ladder case, i.e.,
$r=1,2$.
Hamiltonian in terms of bosons is easily obtained from (\ref{lag2}) and 
(\ref{bosonizedF1}). 
As in the usual cases, it decouples into charge and spin parts.
It is also shown that the gauge fields $A^{(r)}_{\mu}$ appear only
in the form of the Wilson line intergal $\Theta_{r,W}=\int ^{L/2}_{-L/2}dx A^{(r)}_1$
and its conjugate $P_{r,W}$.
\begin{eqnarray}
  H_0
    &\equiv& H_{chain1} + H_{chain2} \nonumber \\
    &=& H_{0C} + H_{0S} \nonumber \\
  H_{0C}
    &=& \displaystyle \sum_{a=\pm} \biggl\{\frac{e^2 L}{2} P_{aW}^2
        + \frac{\pi c}{2L} \Bigl\{ Q_{aC}^2
        + \bigl( Q_{5aC} + \frac{\Theta_{aW}}{\pi} \bigr)^2
        \biggr\} \nonumber \\
    & & {} + \displaystyle \sum_{a=\pm} \int_{-L/2}^{L/2} dx \,
        \frac{c}{2} N_{\sqrt{2} e / \sqrt{\pi c}}
        \bigl( \frac{1}{c^2} \tilde{\Phi}_{aC}^{'2}
        + {\Phi}_{aC}^{'2} + \frac{2e^2}{\pi c}  \Phi_{aC}^2
        \bigr) \label{bosonH0c} \\
  H_{0S}
    &=& \displaystyle \sum_{a=\pm} \frac{\pi c}{2L} \Bigl\{ Q_{aS}^2
        + Q_{5aS}^2 \Bigr\} \nonumber \\
    & & + \displaystyle \sum_{a=\pm}  \int_{-L/2}^{L/2} dx \, \frac{c}{2} N_0
        \biggl( \frac{1}{c^2} \tilde{\Phi}_{aS}^{'2} + {\Phi}_{aS}^{'2}
        \biggr) \label{bosonH0s}
\end{eqnarray}
where $c=aJ/\pi$ and 
\begin{eqnarray}
&& \phi_{r\alpha} ={1 \over \sqrt{2}}(\phi_{rR\alpha}+\phi_{rL\alpha}), \nonumber  \\
&& \tilde{\phi}_{r\alpha} ={1 \over \sqrt{2}}(-\phi_{rR\alpha}+\phi_{rL\alpha}), \nonumber  \\
&& \Phi_{rC}= {1 \over \sqrt{2}}(\phi_{r\uparrow}+\phi_{r\downarrow}),  \nonumber  \\
&& \Phi_{rS}= {1 \over \sqrt{2}}(\phi_{r\uparrow}-\phi_{r\downarrow}),  \nonumber  \\
&& \Phi_{aC}= {1 \over \sqrt{2}}(\Phi_{1C} \pm \Phi_{2C}),  \;\; a=\pm \nonumber  \\
&&  \mbox{similarly for} \; \tilde{\Phi}_{aC}, \Theta_{aW} \; \mbox{etc,}  \nonumber \\
&& Q_{am} = - p_{Lam} + p_{Ram} \nonumber \\
&& Q_{5am} = p_{Lam} + p_{Ram}  \; \; m=S,C \nonumber
\end{eqnarray}
and neutrality condition, which is obtained by varying the action with
respect to $A^{(r)}_0$, is given as 
\begin{eqnarray}
  Q_{+C} &=& L / a = N \nonumber \\
  Q_{-C} &=& 0.
    \label{neutrality condition}
\end{eqnarray}
From (\ref{bosonH0c}), it is obvious that excitations in the charge sector
acquire mass by the gauge interaction and do not include
low-energy modes.

It is convenient to introduce fermions for spin and charge degrees of
freedom,
\begin{eqnarray}
  \psi_{a m \nu}(t,x)
    &=& \frac{1}{\sqrt{L}}
      \, C_{a m \nu} \,
      e^{\pm i \{ q_{a m \nu} + 2\pi p_{a m \nu} (t \pm x)/L \} }
       N_0[ e^{\pm i\sqrt{4\pi}\phi_{a m \nu} (t,x) } ] \nonumber \\
    & & (a=\pm \quad m=S,C \quad \nu=L,R)
  \label{bosonizedF2}
\end{eqnarray}
where
\begin{eqnarray}
  \phi_{\pm m \nu}
    &=& \frac{1}{\sqrt{2}} \bigl(\phi_{1 m \nu} \pm \phi_
        {2 m \nu} \bigl) \\
    \label{boson}
  q_{\pm C \nu}
    &=& \frac{1}{2}\Bigl\{q_{1 \nu \uparrow}+q_{1 \nu \downarrow} \pm
        \bigl(q_{2 \nu \uparrow}+q_{2 \nu \downarrow} \bigl) \Bigl\}
        \nonumber \\
  q_{\pm S \nu}
    &=& \frac{1}{2}\Bigl\{q_{1 \nu \uparrow}-q_{1 \nu \downarrow} \pm
        \bigl(q_{2 \nu \uparrow}-q_{2 \nu \downarrow} \bigl) \Bigl\}
  \label{zeromode}
\end{eqnarray}
likewise for $p$ and 
\begin{eqnarray}
  C_{\pm C \nu}
    &=& \Bigl\{C_{1 \nu \uparrow} C_{1 \nu \downarrow} \bigl(C_{2 \nu
    \uparrow} C_{2 \nu \downarrow} \bigl)^{\pm 1} \Bigl\}^{1/2} \nonumber \\
  C_{\pm S \nu}
    &=& \Bigl\{C_{1 \nu \uparrow} C_{1 \nu \downarrow}^{-1} \bigl(C_{2 \nu
        \uparrow} C_{2 \nu \downarrow}^{-1} \bigl)^{\pm 1} \Bigl\}^{1/2}.
  \label{Csc}
\end{eqnarray}
In Appendix A, $C$'s are explicitly given, and with them it is verified that
$\psi_{am\nu}$'s satisfy the CAR.
Then it is shown that spin-part Hamiltonian (\ref{bosonH0s}) is nothing but 
that of free massless Dirac fields $\psi_{\pm S}=(\psi_{\pm SR}\psi_{\pm SL})$. 

Let us consider the inter-chain interaction.
Continuum limit of the spin operators $\vec{S}_{i,r}$ are given in terms of
low-energy fermions as follows,
\begin{equation}
\vec{S}_{i,r} \rightarrow l\vec{S}_r(x), \;\; 
\vec{S}_r(x)=\vec{J}_{rR}(x)+\vec{J}_{rL}(x)+(-1)^i\vec{n}_r(x),
\label{conS}
\end{equation}
where smooth part of the $SU(2)$ currents are given as 
\begin{equation}
\vec{J}_{r\nu}=:\psi^{\dagger}_{r\nu}{\vec{\sigma} \over 2}\psi_{r\nu}:,
\label{smoothS}
\end{equation}
and staggered part is 
\begin{equation}
\vec{n}_r(x)=:\psi^{\dagger}_{rL}{\vec{\sigma} \over 2}\psi_{rR}: +
\; \mbox{h.c.}
\label{staggeredS}
\end{equation}
In terms of these operators, the inter-rung Hamiltonian is expressed as
\begin{equation}
H_{rung}=J'l \int dx \Big[ \vec{J}_1(x)\cdot \vec{J}_2(x)+
\vec{n}_1(x)\cdot \vec{n}_2(x)\Big].
\label{Hrung}
\end{equation}
From (\ref{bosonizedF1}) and (\ref{bosonH0c}), it is easily seen that the $\vec{J}$-term
in (\ref{Hrung}) is marginal whereas the $\vec{n}$-term is relevant.
Therefore we shall consider effects of the $\vec{n}$-term first.
In term of the fermions $\psi_{am\nu}$ defined by (\ref{bosonizedF2}), 
the relevant interaction term
is written as 
\begin{eqnarray}
H^{rel}_{rung} &=& J'l \int dx \; \vec{n}_1(x)\cdot \vec{n}_2(x)  \nonumber  \\
&=& {J'l \over 4}\int  dx \Big(\psi^{\dagger}_{+CL}\psi_{+CR}
+\psi^{\dagger}_{-CL}\psi_{-CR}- \mbox{h.c.}\Big)  \nonumber  \\
&& \times \Big(\psi^{\dagger}_{+SL}\psi_{+SR}-\psi^{\dagger}_{-SL}\psi_{-SR}
-2\psi^{\dagger}_{-SL}\psi^{\dagger}_{-SR}-  \mbox{h.c.}\Big),
\label{Hrel}
\end{eqnarray} 
and the Hamiltonian under consideration is given by
\begin{equation}
H^{rel}_2=H_{0C}+H_{0S}+H^{rel}_{rung}.
\label{Hr2}
\end{equation}
By introducing auxilary fields or mean-field type calculation for the four-Fermi
interaction $H^{rel}_{rung}$, we can determine the mass of $\psi_{am\nu}$.
Action with auxiliary fields is explicitly given as
\begin{eqnarray}
S[\psi, \sigma] &=& S_0[\psi] +i{J'l \over 4}\int dx\; \sigma
\Big( \psi^{\dagger}_{+CL}\psi_{+CR}
+\psi^{\dagger}_{-CL}\psi_{-CR}- \mbox{h.c.}\Big)   \nonumber   \\
&& +i{J'l \over 4}\int dx\; \sigma' \Big(\psi^{\dagger}_{+SL}\psi_{+SR}-\psi^{\dagger}_{-SL}\psi_{-SR}
-2\psi^{\dagger}_{-SL}\psi^{\dagger}_{-SR}-  \mbox{h.c.}\Big)  \nonumber   \\
&& +{J'l \over 4}\int dx \; \sigma\sigma',
\label{Saux}
\end{eqnarray}
where $S_0[\psi]$ is action for the Hamiltonian $H_{0C}+H_{0S}$.
Gap equations which determine values of $\sigma$ and $\sigma'$ are
given as 
\begin{eqnarray}
&& \langle \psi^{\dagger}_{+CL}\psi_{+CR}
+\psi^{\dagger}_{-CL}\psi_{-CR}- \mbox{h.c.} \rangle_{\sigma_0}
=i\sigma'_0,   \label{gapeq1}   \\
&& \langle \psi^{\dagger}_{+SL}\psi_{+SR}-\psi^{\dagger}_{-SL}\psi_{-SR}
-2\psi^{\dagger}_{-SL}\psi^{\dagger}_{-SR}-  \mbox{h.c.} \rangle_{\sigma'_0}
=i \sigma_0,
\label{gapeq2}
\end{eqnarray}
where $\langle \cdot\cdot\cdot \rangle_{\sigma_0}$ etc. denote expectation value
evaluated with putting $\sigma=\sigma_0$ etc. in $S[\psi, \sigma]$ in (\ref{Saux}).
Evaluation of the LHS on (\ref{gapeq2}) is straightforward in terms of {\em Majorana fermion}
representation of $H_{0S}$.
Majorana fermions are defined as 
\begin{eqnarray}
&& \xi^1_{\nu}={ 1 \over \sqrt{2}}(\psi_{+S\nu}+\psi^{\dagger}_{+S\nu}), \;\;
  \xi^2_{\nu}={ 1 \over \sqrt{2}i}(\psi_{+S\nu}-\psi^{\dagger}_{+S\nu}),   \nonumber \\
&& \xi^3_{\nu}={ 1 \over \sqrt{2}}(\psi_{-S\nu}+\psi^{\dagger}_{-S\nu}), \;\;
\rho_{\nu}={ 1 \over \sqrt{2}i}(\psi_{+S\nu}-\psi^{\dagger}_{+S\nu}), \;\; (\nu =R,L)
\label{majorana}
\end{eqnarray}
and 
\begin{eqnarray}
\sigma_0&=& -i \Big\{3\langle \rho_L\rho_R\rangle -\sum_{n=1,2,3}\langle \xi^n_L\xi^n_R
\rangle \Big\}  \nonumber  \\
&=& -{3 \over 2\pi}{M_s \over c} \Big\{3 \ln \Big({\Lambda c \over 3M_s}
 +\sqrt{1+({\Lambda c \over 3M_s})^2} \Big)  \nonumber  \\
 && \;\; +\ln \Big({\Lambda c \over M_s}
 +\sqrt{1+({\Lambda c \over M_s})^2} \Big) \Big\},
\label{Mc}
\end{eqnarray}
where $M_s={J'l \over 4}\sigma'_0$.

On the other hand for the charge sector, there already exists mass term
of the boson fields $\Phi_{aC}$ as seen from (\ref{bosonH0c}).
Therefore calculation of the LHS on (\ref{gapeq1}) needs some approximation.
Here we employ perturbative calculation in powers of
 $\sigma_0$ assuming that the boson
mass term $\Phi^2$ is more relevent than the $\cos \sqrt{4\pi} \Phi$ 
term.\footnote{This is certainly correct as seen from the conformal dimensions of these
operators.
For renormalization-group study of the massive sine-Gordon theory,
see Ref.\cite{RGSG}.}
In the study of the  Schwinger model, vacuum expectation values (VEV's)
similar to the above operators are evaluated\cite{HH}.
Here we use similar argument.

From the constraints (\ref{neutrality condition}), the Hamiltonian $H_{0C}$
is given as 
\begin{eqnarray}
H_{0C}&=& {e^2 L \over 2}P^2_{+W}+{\pi c \over 2L}\Big\{N^2
+(2p_{R+C}+N+{\Theta_{+W} \over \pi})^2 \Big\} \nonumber   \\
&& +{e^2L \over 2}P^2_{-W} +{\pi c \over 2L}+{\pi c \over 2L}(2p_{R-C}
+{\Theta_{+W} \over \pi})^2 + \; \mbox{nonzero modes}.
\label{H0c2}
\end{eqnarray}
Zero-mode part of the wave function of the above  Hamiltonian
is labeled by integers $(p_{R+C}=n_+, p_{R-C}= n_-)$ and wave function is 
explicitly given as
\begin{equation}
\langle q_+,q_- | n_+,n_- \rangle = e^{-i(q_+n_++q_-n_-)}.
\label{wavefunc}
\end{equation}
If there is no mass term of fermions $\psi_{\pm C\nu}$, the above states
$| n_+,n_- \rangle$  are degenerate.
The mass term, 
\begin{equation}
-iM_c(\psi^{\dagger}_{+CL}\psi_{+CR}+\psi^{\dagger}_{-CL}\psi_{-CR}
- \mbox{h.c.}), \; \; M_c={J'l \over 4}\sigma_0,
\label{massterm}
\end{equation}
has nonvanishing matrix element between the states with
different $(n_+, n_-)$:
\begin{equation}
\langle l_+, l_- |\psi^{\dagger}_{+CL}\psi_{+CR}|n_+, n_- \rangle 
\propto \delta_{l_++1,n_+}\delta_{l_-,n_-},
\label{tunnel} 
\end{equation}
etc.
Then we should define the $\theta$-vacuum by taking linear combination of them,
\begin{equation}
| \theta_+, \theta_-\rangle \equiv \sum_{n_+,n_-}
e^{i\theta_+n_++i\theta_-n_-}|n_+, n_- \rangle.
\label{thetav}
\end{equation}
Including the contribution from the nonzero modes, we evaluate the matrix
element in the $\theta$-vacuum as follows,
\begin{equation}
\langle \psi^{\dagger}_{+CL}\psi_{+CR} \rangle _{\theta}=
e^{-i\theta_+}e^{\gamma}\sqrt{{2e^2 \over \pi c}},
\label{VEV}
\end{equation}
etc., where $\gamma$ is the Euler constant.
Then the VEV of the mass term (\ref{massterm}) depends on the values of
the $\theta$'s.
It has the largest value for $\theta_+=\theta_-=\pm \pi /2$, at which 
the ground state energy of the system has the lowest value.
The final result is
\begin{equation}
M_s=\pm {J'l \over 4\pi}e^{\gamma}\sqrt{{2e^2 \over \pi c}}.
\label{Ms}
\end{equation}

Low-energy Hamiltonian of the spin sector is then given by 
\begin{eqnarray}
H_s &=& H_{0S}+H_{rung;S}  \nonumber   \\
&=& \int ^{L/2}_{-L/2}dx \Big[ \sum_{n=1,2,3}\Big\{{ic \over 2}(\xi^n_L\partial_x
\xi^n_L-\xi^n_R\partial_x \xi^n_R)+iM_s\xi^n_L\xi^n_R\Big\}  \nonumber   \\
&& \; \; +\Big\{{ic \over 2}(\rho_L\partial_x\rho_L-\rho_R\partial_x\rho_R)
-3iM_s\rho_L\rho_R\Big\}\Big].
\label{Hs}
\end{eqnarray}
The same Hamiltonian with the above was obtained by Shelton et.al.\cite{shelton}
from the Hubbard Hamiltonian, but as we mentioned in the introduction,
straightforward use of the bosonization formulas {\em ignoring the zero modes}
does not give (\ref{Hs}).
From the derivation in this paper, it is obvious that the zero modes in the
bosonization of fermions play an important role and we have to consider
the $\theta$-vacuum in order to have the correct ground state of the system.

From (\ref{Hs}), low-energy excitations in the 2-leg quantum spin
ladder are described by spin-triplet
Majorana fermions $\xi^n_{\nu}$ and spin singlet fermion $\rho_{\nu}$.
This was verified by experiments\cite{kishine}.
Marginally relevant interactions like $\vec{J}_1\cdot \vec{J}_2$
give only renormalization of velocity and mass $M_s$.

Recently Hosotani studied quantum spin ladder by a similar method with ours,
but he did not obtain the above result\cite{hosotani}.
In his approach, nearest-neighbor(NN) fermions in the {\em real space} are
paired in order to obtain relativistic Dirac fermions, though we have paired
right-moving and left-moving fermions near the Fermi points in the momentum
space, as it should be.
We are not quite certain if real-space paring gives correct continuum limit
which describes low-energy excitations.
We also suspect that the bosonized expression of inter-chain interaction
is not correct.
This may be connected with the above problem.
The inter-chain interaction should contain not only the scalar fields $\phi_{\alpha}$
but also the dual scalar fields $\tilde{\phi}_{\alpha}$
because of the term $(S^+_1S^-_2 +\mbox{h.c.})$ in the Heisenberg Hamiltonian, 
but his expression does not.
Because of this, his result does not seem to be SU(2) invariant.

\setcounter{equation}{0}
\section{AF Heisenberg model on 3-leg ladder}

In the previous section, we studied low-energy excitations in 2-leg spin ladder.
In this section, we shall apply similar analysis to the 3-leg case in which
excitations are expected as gapless.
Hamiltonian is given by (\ref{Hl}) with $l_c=3$, and the same gauge-theoretical
description is possible for each chain.
Most relevent inter-chain interaction is described by a similar Hamiltonian
to the $\vec{n}$-term in (\ref{Hrung}), 
\begin{equation}
H^{rel}_{3rung}=J'l \int dx \; \Big( \vec{n}_1(x)\cdot \vec{n}_2(x)+
\vec{n}_2(x)\cdot \vec{n}_3(x)\Big).
\label{Hrel3}
\end{equation}
Then it is convenient to introduce the following boson fields which describe
excitations in the  (1-2) chains, etc.,
\begin{eqnarray}
  \varphi_{\pm A \nu}
    &=& \bigl( \phi_{1 \nu} \pm \phi_{2 \nu} \bigr) / \sqrt{2},
      \nonumber \\
  \varphi_{\pm B \nu}
    &=& \bigl( \phi_{3 \nu} \pm \phi_{2 \nu} \bigr) / \sqrt{2},
      \label{3boson} \\
  \varphi_{\pm C \nu}
    &=& \bigl( \phi_{3 \nu} \pm \phi_{1 \nu} \bigr) / \sqrt{2}, \;\; (\nu=R,L)
      \nonumber
\end{eqnarray}
where the zero modes are defined in the same way.
We shall consider only spin excitations and  have omitted subscript ``$S$",
i.e., $\phi_{r}=(\phi_{r\uparrow}
-\phi_{r\downarrow})/\sqrt{2}$ $(r=1,2,3)$ in (\ref{3boson}) since charge
excitations are massive as in the 2-leg case,
$$
e^2 \sum_{i=1,2,3}\Phi^2_{iC} \propto e^2 \sum_{X=A,B,C}\Phi^2_{\pm XC},
$$
where notations are self-evident.
It is also useful to introduce fermions $\Psi_{\pm X}$ $(X=A,B,C)$
corresponding to the above boson fields for each channel.
As we showed in the previous section, the fermions $\Psi_{\pm A,B}$ become
massive as a result of the inter-chain interactions $H^{rel}_{3rung}$.
Spin part of the Hamiltonian of the 3-leg spin ladder is then given by
\begin{equation}
H^{rel}_{3S}=\sum_{X=A,B,C}\Big( H_{+X}+H_{-X}\Big),
\label{Hrel3s}
\end{equation} 
where
\begin{eqnarray}
  H_{+ X}
    &=& \int dx \, \Bigl\{i \frac{c}{2} \bigl(
        \Psi_{+ X L}^{\dagger} \partial_x \Psi_{+ X L}
        - \Psi_{+ X R}^{\dagger} \partial_x \Psi_{+ X R}
        \bigr) \nonumber \\
    & & {} + i M_s \bigl(1-\delta_{X,C} \bigr)
        \bigl( \Psi_{+ X L}^{\dagger} \Psi_{+ X R}
        - \mbox{h.c.} \bigr) \Bigr\}
        \label{eq:H+} \\
  H_{- X}
    &=& \int dx \, \Bigl\{i \frac{c}{2} \bigl(
        \Psi_{- X L}^{\dagger} \partial_x \Psi_{- X L}
        - \Psi_{- X R}^{\dagger} \partial_x \Psi_{- X R}
        \bigr) \nonumber \\
    & & {} - i M_s \bigl(1-\delta_{X,C} \bigr)
        \bigl( \Psi_{- X L}^{\dagger} \Psi_{- X R}
        - \mbox{h.c.} \bigr) \nonumber \\
    & & {} - 2i M_s \bigl(1-\delta_{X,C} \bigr)
        \bigl( \Psi_{- X L}^{\dagger} \Psi_{- X R}^{\dagger}
        - \mbox{h.c.} \bigr) \Bigr\}.
        \label{eq:H-}
\end{eqnarray}
The $A,B$ part of the above Hamiltonian is the same with that in the
2-leg case and $M_s=\pm (J' l /4 \pi)e^\gamma \sqrt{2 e^2 / \pi c}$,
where we set $\theta_{\pm A} \equiv \theta_1 \pm \theta_2=
\theta_{\pm B} \equiv \theta_3 \pm \theta_2=\pm {\pi \over 2}$.
The $H_{\pm C}$ is the system of massless fermions (i.e., $H_{0S}$).
In the above expression (\ref{Hrel3s}), (\ref{eq:H+}) and (\ref{eq:H-}),
we have neglected the marginally relevant interactions of fermions
which appear from $\vec{J}$-terms in the inter-chain interaction.
These terms cannot be neglected in the present 3-leg system
and we shall discuss it later on.

From (\ref{3boson}), it is obvious that not all the bosons $\varphi_{\pm X}$
or fermions $\Psi_{\pm X}$ are independent.
We have to impose the following constraints: 
\begin{eqnarray}
  \varphi_{+ A \nu} - \varphi_{- A \nu}
    &=& \varphi_{+ B \nu} - \varphi_{- B \nu},    \nonumber \\
  \varphi_{+ B \nu} + \varphi_{- B \nu}
    &=& \varphi_{+ C \nu} + \varphi_{- C \nu},
        \label{const1} \\
  \varphi_{+ C \nu} - \varphi_{- C \nu}
    &=& \varphi_{+ A \nu} + \varphi_{- A \nu}, \;\; (\nu=R,L)  \nonumber
\end{eqnarray}
and similarly for zero modes.
These constraints can be rewritten in terms of fermions by using the
following identities,
\begin{eqnarray}
  \Psi_{\pm X L}^{\dagger} \Psi_{\pm X L}
    &=& p_{\pm X L} / L
       + \partial_x \varphi_{\pm X L} / \sqrt{\pi},
       \nonumber \\
   \Psi_{\pm X R}^{\dagger} \Psi_{\pm X R}
    &=& p_{\pm X R} / L
       + \partial_x \varphi_{\pm X R} / \sqrt{\pi}.
       \label{const2}
\end{eqnarray}
From (\ref{const1}) and (\ref{const2}),\footnote{Here we neglect the
constraints on the $q$-terms in the bosonization.
This is justified, since operator $q$ is the conjugate to the momentum $p$ and
wave function is given as an eigen state of $p$ as discussed in Sect.2.}
\begin{eqnarray}
  \Psi_{+ A \nu}^{\dagger} \Psi_{+ A \nu}
       - \Psi_{- A \nu}^{\dagger} \Psi_{- A \nu}
    &=& {} \Psi_{+ B \nu}^{\dagger} \Psi_{+ B \nu}
       - \Psi_{- B \nu}^{\dagger} \Psi_{- B \nu},    \nonumber \\
  \Psi_{+ B \nu}^{\dagger} \Psi_{+ B \nu}
       + \Psi_{- B \nu}^{\dagger} \Psi_{- B \nu}
    &=& {} \Psi_{+ C \nu}^{\dagger} \Psi_{+ C \nu}
       + \Psi_{- C \nu}^{\dagger} \Psi_{- C \nu},
       \label{const3} \\
  \Psi_{+ C \nu}^{\dagger} \Psi_{+ C \nu}
       - \Psi_{- C \nu}^{\dagger} \Psi_{- C \nu}
    &=& {} \Psi_{+ A \nu}^{\dagger} \Psi_{+ A \nu}
       + \Psi_{- A \nu}^{\dagger} \Psi_{- A \nu}. \nonumber
\end{eqnarray}
The above constraints (\ref{const3}) can be imposed by Lagrange multiplyers
$\lambda_{i\nu}$ $(i=1,2,3)$ and we add the following terms to the 
Hamiltonian $H^{rel}_{3S}$ (\ref{Hrel3s}),
\begin{eqnarray}
   \lefteqn{ \lambda_{1 \nu} \bigl(\Psi_{+ A \nu}^{\dagger}
    \Psi_{+ A \nu}- \Psi_{- A \nu}^{\dagger} \Psi_{- A \nu}
       - \Psi_{+ B \nu}^{\dagger} \Psi_{+ B \nu}
       + \Psi_{- B \nu}^{\dagger} \Psi_{- B \nu}
        \bigr)}\hspace{0.5cm} \nonumber \\
   \lefteqn{\  + \lambda_{2 \nu} \bigl(
        \Psi_{+ B \nu}^{\dagger} \Psi_{+ B \nu}
       + \Psi_{- B \nu}^{\dagger} \Psi_{- B \nu}
       - \Psi_{+ C \nu}^{\dagger} \Psi_{+ C \nu}
       - \Psi_{- C \nu}^{\dagger} \Psi_{- C \nu}
       \bigr)}\hspace{0.3cm} \nonumber \\
   \lefteqn{\  + \lambda_{3 \nu} \bigl(
       \Psi_{+ C \nu}^{\dagger} \Psi_{+ C \nu}
       - \Psi_{- C \nu}^{\dagger} \Psi_{- C \nu}
       - \Psi_{+ A \nu}^{\dagger} \Psi_{+ A \nu}
       - \Psi_{- A \nu}^{\dagger} \Psi_{- A \nu}
       \bigr)}\hspace{0cm} \nonumber \\
    &=& (\lambda_{1 \nu} - \lambda_{3 \nu})
        \Psi_{+ A \nu}^{\dagger} \Psi_{+ A \nu}
        - (\lambda_{1 \nu} + \lambda_{3 \nu})
        \Psi_{- A \nu}^{\dagger} \Psi_{- A \nu}
        \nonumber \\
    & & {} + (\lambda_{2 \nu} - \lambda_{1 \nu})
        \Psi_{+ B \nu}^{\dagger} \Psi_{+ B \nu}
        + (\lambda_{2 \nu} + \lambda_{1 \nu})
        \Psi_{- B \nu}^{\dagger} \Psi_{- B \nu}
        \nonumber \\
    & & {} \quad \! + (\lambda_{3 \nu} - \lambda_{2 \nu})
        \Psi_{+ C \nu}^{\dagger} \Psi_{+ C \nu}
        - (\lambda_{3 \nu} + \lambda_{2 \nu})
        \Psi_{- C \nu}^{\dagger} \Psi_{- C \nu}
        \label{eq:3Lm}.
\end{eqnarray}
Through the above terms, $A, B$ and $C$ fermions are interacting
with each other.

We are interested in how the constraints (\ref{const3}) influence 
low-energy excitations, i.e., the $C$ fermions which are gapless
if we ignore the constraints.
To this end, we shall use a similar argument as  in the study of the
nonlinear-$\sigma$ model, i.e., we integrate out the massive fermions
$\Psi_{\pm A\nu}$ and $\Psi_{\pm B\nu}$ and see whether mass terms of 
the Lagrange multiplyers appear or not.
If mass terms of $\lambda_{i\nu}$ are generated, the constraints (\ref{const3})
tend to irrelevant for low-energy excitations, whereas if $\lambda_{i\nu}$ remains
massless the constraint is intact. 

The action of the $+A$-sector $S_{+A}$ is written as
\begin{eqnarray}
  S_{+A}
    &=& \int d^2 x \,  \Bigl[ \bar{\Psi}_{+A}
           \gamma_{\mu} \bigl(  \partial_{\mu} +
           i \tilde{A}_{\mu} \bigr) \Psi_{+A}
          + i (2 M_s/c) \bar{\Psi}_{+A}\gamma_5
          \Psi_{+A} \bigr]
          \label{S+A}
\end{eqnarray}
where we put $x_0 = c \tau / 2$ and 
\begin{eqnarray}
  \tilde{A}_0
    &=& -i \bigl( \lambda_{1L} - \lambda_{3L} +
           \lambda_{1R} - \lambda_{3R} \bigr) / 2,
           \nonumber \\
  \tilde{A}_1
    &=& - \bigl( \lambda_{1L} - \lambda_{3L} - \lambda_{1R}
          + \lambda_{3R} \bigr) / 2.
          \nonumber
\end{eqnarray}
Then effective potential of the $\lambda_{i\nu}$ is obtained as\footnote{Please remark
that regularization scheme is specified by the spatial momentum cutoff $\Lambda$.}
\begin{eqnarray}
  - S_{+A}^{eff} [\lambda]
    &=& -i \int d^2 x \, \bigl\langle \bar{\Psi}_{+A}(x)
          \gamma_{\mu} A_{\mu} \Psi_{+A}(x)
         \bigr\rangle \nonumber \\
    & & - \frac{1}{2} \int d^2 x d^2 x' \, \bigl\langle\bar{\Psi}_{+A}(x)
       \gamma_{\mu} A_{\mu} \Psi_{+A}(x)
          \bar{\Psi}_{+A}(x') \gamma_{\mu} A_{\mu} \Psi_{+A}(x')
          \bigr\rangle _c
          \nonumber \\
   &=& i \int d^2 x \int \frac{d^2 p} {(2 \pi)^2}
          Tr \bigl[ g_+ (p) \gamma_{\mu} A_{\mu} \bigr]
          \nonumber \\
   & & {} + \frac{1}{2} \int d^2 x \int \frac{d^2 p} {(2 \pi)^2}
        Tr \bigl[ g_+ (p) \gamma_{\mu} A_{\mu}
        g_+ (p) \gamma_{\mu} A_{\mu}
       \bigr] \nonumber \\
   &=& \int d^2 x \, a_P
        \bigl(  \lambda_{1L} - \lambda_{3L} - \lambda_{1R}
        + \lambda_{3R} \bigr) ^2
       \label{eS+A}
\end{eqnarray}
where
\begin{eqnarray}
  g_+ (p)
    &=& \bigl( \gamma_{\mu} \partial_{\mu} +i (2 M_S / c)
           \gamma_5 \bigr)^{-1} (p)
          \nonumber \\
    &=& - \frac{i p_{\mu} \gamma_{\mu} + i (2 M_S / c)
           \gamma_5}{p^2 + (2M_S/c)^2},
          \label{eq:g+}  \\
  a_P &=& \frac{1}{8 \pi} \frac{\Lambda c/2 M_S}
            {\sqrt{1+(\Lambda c/2 M_S)^2}}.
          \label{eq:consta}
\end{eqnarray}
The action of the $-A$-sector is given in terms of Majorana fermions defined by
\begin{eqnarray}
\zeta^3_{\nu} &=& ( \Psi_{-A \nu} + \Psi_{-A \nu}^{\dagger})/2¡¤\nonumber  \\
\chi_{\nu} &=& ( \Psi_{-A \nu} - \Psi_{-A \nu}^{\dagger})/2i,
\label{majorana2}
\end{eqnarray}
and
\begin{eqnarray}
  S_{-A}
    &=& \int d^2 x \, \bigl[ \frac{1}{2} \bar{\zeta}^3 
           \bigl( \gamma_{\mu} \partial_{\mu} +
           i (2 M_S /c) \gamma_5 \bigr) \zeta^3
           \nonumber \\
    & &  {} + \frac{1}{2} \bar{\chi} 
           \bigl( \gamma_{\mu} \partial_{\mu} -
           i (6 M_S /c) \gamma_5 \bigr) \chi
           \nonumber \\
    & & {} + i \bar{\zeta}^3 \gamma_{\mu} \tilde{B}_{\mu}
          \chi \bigr] ,
\label{S-A}
\end{eqnarray}
where
\begin{eqnarray}
  \tilde{B}_0
    &=& \bigl( \lambda_{1L} + \lambda_{3L} + \lambda_{1R}
          + \lambda_{3R} \bigr)/2,
          \nonumber \\
  \tilde{B}_1
     &=& -i \bigl( \lambda_{1L} + \lambda_{3L} - \lambda_{1R}
           - \lambda_{3R} \bigr)/2.
          \nonumber
\end{eqnarray}
Then contribution from the $-A$-section to the effective potential of the 
$\lambda_{i\nu}$ is obtained as
\begin{eqnarray}
  -S_{-A}^{eff} [\lambda]
    &=&  -i \int d^2 x \, \bigl\langle \bar{\zeta}^3(x)
          \gamma_{\mu} A_{\mu} \chi (x)
         \bigr\rangle \nonumber \\
    & & - \frac{1}{2} \int d^2 x d^2 x' \, \bigl\langle\bar{\zeta}^3(x)
   \gamma_{\mu} A_{\mu} \chi (x)
          \bar{\zeta}^{3}(x') \gamma_{\mu} A_{\mu} \chi(x')
          \bigr\rangle _c
          \nonumber \\
   &=& - \frac{1}{2} \int d^2 x \int \frac{d^2 p} {(2 \pi)^2}
        Tr \bigl[ g_1 (p) \gamma_{\mu} A_{\mu}
        g_2 (p) \gamma_{\mu} A_{\mu}
       \bigr] \nonumber \\
   &=& \int d^2 x \, \Bigl[ b_P \bigl\{ \bigl( \lambda_{1L}
          + \lambda_{3L} \bigr)^2 + \bigl( \lambda_{1R}
          + \lambda_{3R} \bigr)^2 \bigr\}
          \nonumber \\
    & & {} +c_P \bigl(\lambda_{1L} + \lambda_{3L} \bigr)
          \bigl(\lambda_{1R} + \lambda_{3R} \bigr) \Big],
       \label{eS-A}
\end{eqnarray}
where
\begin{eqnarray}
  g_1(p)
    &=& \bigl(\gamma_{\mu} \partial_{\mu} + i(2M_S /c)
           \gamma_5 \bigr)^{-1} (p)
          \nonumber \\
    &=& - \frac{i p_{\mu} \gamma_{\mu} + i (2 M_S / c)
           \gamma_5}{p^2 + (2M_S/c)^2}
          \label{eq:g1}  \\
   g_2(p)
    &=& \bigl(\gamma_{\mu} \partial_{\mu} - i(6M_S /c)
           \gamma_5 \bigr)^{-1} (p) \nonumber \\
    &=& - \frac{i p_{\mu} \gamma_{\mu} - i (6 M_S / c)
           \gamma_5}{p^2 + (6M_S/c)^2}
          \label{eq:g2}  \\
  b_P
    &=& \frac{1}{32 \pi} \frac{\Lambda c}{2 M_S}
           \bigl(\sqrt{9+(\Lambda c/2 M_S)^2} -
           \sqrt{1+(\Lambda c /2M_S)^2} \bigr)
           \label{eq:constb} \\
  c_P
    &=& \frac{3}{16 \pi} \ln \biggl( \frac{ \Lambda c/2 M_S
           + \sqrt{1+(\Lambda c/2M_S)^2}} {\Lambda c/6 M_S
           + \sqrt{1+(\Lambda c/6M_S)^2}} \biggr).
           \label{constc}
\end{eqnarray}
Similarly contributions from $+B$, $-B$-sectors are evaluated as 
\begin{eqnarray}
  - S_{+B}^{eff} [\lambda]
   &=& \int d^2 x \, a_P
        \bigl(  \lambda_{2L} - \lambda_{1L} - \lambda_{2R}
        + \lambda_{1R} \bigr)^2,
       \label{eq:e+SB} \\
   -S_{-B}^{eff} [\lambda]
   &=& \int d^2 x \, \Bigl[ b_P \bigl\{ \bigl( \lambda_{2L}
          + \lambda_{1L} \bigr)^2 + \bigl( \lambda_{2R}
          + \lambda_{1R} \bigr)^2 \bigr\}
          \nonumber \\
    & & {} +c_P \bigl(\lambda_{2L} + \lambda_{1L} \bigr)
          \bigl(\lambda_{2R} + \lambda_{1R} \bigr) \Big].
       \label{e-SB}
\end{eqnarray}
Then the potential is given as 
\begin{eqnarray}
  -S^{eff}
   &=& S_{+A}^{eff} + S_{-A}^{eff} + S_{+B}^{eff} + S_{-B}^{eff}
          \nonumber \\
    &=&\int d^2x \Big[ 2(a_P+b_P) \lambda_1^2 + (a_P+b_P) \lambda_2^2
          + (a_P+b_P) \lambda_3^2
          \nonumber \\
    & & {} -(2a_P-c_P) \lambda_1 \Gamma_0
           \lambda_1 - (a_P-c_P/2) \lambda_2 \Gamma_0 \lambda_2
         - (a_P-c_P/2) \lambda_3 \Gamma_0 \lambda_3
           \nonumber \\
    & & {}  -2 (a_P-b_P) \lambda_1 (\lambda_2 + \lambda_3 )
          +(2a_P+c_P) \lambda_1 \Gamma_0 ( \lambda_2 + \lambda_3 )\Big],
\label{totalpot}    
\end{eqnarray}
where $\lambda_i=(\lambda_{iR} \lambda_{iL})$ and $\Gamma_0=\sigma_1$.

As we mentioned before, there are also contributions from marginally
relevent four-Fermi interactions which are generated by the smooth part
of the spin interactions, renormalization effect of high-momentum modes, etc.
We first ignore their effects.
Then we intergate over the field $\lambda_1$ in $S^{eff}$ and obtain
\begin{eqnarray}
  -S'^{eff}[ \lambda_2 , \lambda_3 ]
    &=&\int d^2x \Big[ \frac{1}{2} \bigl( \lambda_2 + \lambda_3 \bigr)
           \bigl( \eta - \omega \Gamma_0 \bigr)
           \bigl( \lambda_2 + \lambda_3 \bigr)
           \nonumber \\
     & & {} + \frac{1}{2}\bigl( \lambda_2 - \lambda_3
           \bigr) \bigl( 1- \alpha \Gamma_0 \bigr) \bigl(
           \lambda_2 - \lambda_3 \bigr)\Big],
            \label{effS'}
\end{eqnarray}
where
\begin{eqnarray}
  \alpha &=& \bigl( a_P -c_P/2 \bigr) / \bigl( a_P+b_P \bigr)
                    \nonumber \\
  \beta &=& \sqrt{2} \bigl( a_P -b_P \bigr) / \bigl( a_P +b_P \bigr)
                   \nonumber \\
  \delta &=& \sqrt{2} \bigl( a_P +c_P/2 \bigr) / \bigl( a_P +b_P \bigl)
                    \nonumber  \\
  \eta &=& 1 + \bigl( \beta^2 + \delta^2 - 2 \alpha \beta \delta
                 \bigr) / 2 \bigl(1- \alpha^2 \bigr) \nonumber \\
  \omega &=& \alpha - \bigl\{ \alpha ( \beta^2 + \delta^2 )
                   - 2 \beta \delta \bigr\} / 2 \bigl( 1-\alpha^2 \bigr).
                  \label{coeff2}
\end{eqnarray}
Then action of the $C$-sector is obtained as follows:
\begin{eqnarray}
  S_{+C}
    &=&\int d^2x \Big[ \bar{\Psi}_{+C} \gamma_{\mu} \partial_{\mu} \Psi_{+C}
           + \bigl( \lambda_{2 \nu} - \lambda_{3 \nu} \bigr)
           \Psi_{+C \nu}^{\dagger} \Psi_{+C \nu}
           \nonumber \\
    & & {} - \bigl( \lambda_2 - \lambda_3 \bigr) \bigl( 1- \alpha
          \Gamma_0 \bigr) \bigl( \lambda_2 - \lambda_3 \bigr)\Big],
           \label{effS+C} \\
  S_{-C}
    &=&\int d^2x \Big[ \bar{\Psi}_{-C} \gamma_{\mu} \partial_{\mu} \Psi_{-C}
           + \bigl( \lambda_{2 \nu} + \lambda_{3 \nu} \bigr)
           \Psi_{-C \nu}^{\dagger} \Psi_{-C \nu}
           \nonumber \\
    & & {} - \frac{1}{2} \bigl( \lambda_2 + \lambda_3 \bigr)
           \bigl( \eta - \omega \Gamma_0 \bigr)
           \bigl( \lambda_2 + \lambda_3 \bigr)\Big].
           \label{effS-C}
\end{eqnarray}
By straightforward calculation, it is verfied that the mass matrix of the
fields $\lambda_{2,3}$ in (\ref{effS+C}) and  (\ref{effS-C}) has positive eigen-values for the
coefficients (\ref{coeff2}), and then $\lambda_{2,3}$ can be integrated out.
Marginally relevant terms appear like $\Psi^{\dagger}_{+CR}\Psi_{+CR}
\Psi^{\dagger}_{+CL}\Psi_{+CL}$ but the $C$-sector still remains gapless.
The investigation given so far shows that spin excitations on the top and
bottom chains are gapless and described by two massless boson fields
$\varphi_{\pm C\nu}$, though excitations in the middle chain have a gap.

We shall comment on effect of the marginal interactions which we have
ignored so far.
As in the 2-leg case, the SU(2) current $I^l_{\nu}$ is given
as follows in terms of the spin-triplet Majorana fermions,
\begin{equation}
\zeta^1_{\nu}=(\Psi_{+A\nu}+\Psi^{\dagger}_{+A\nu})/2,  \;\;
\zeta^2_{\nu}=(\Psi_{+A\nu}-\Psi^{\dagger}_{+A\nu})/2i,
\nonumber 
\end{equation}
\begin{equation}
I^l_{A\nu}=i\epsilon^{lmn}\zeta^m_{\nu}\zeta^m_{\nu},
\label{current2}
\end{equation}
where $\epsilon^{lmn}$ is the antisymmetric tensor.
Then interaction terms are given as
\begin{equation}
H^{mar}_{A}=\int dx \Big[g_1\sum_l I^l_{AR}I^l_{AL} + g_2(\sum_{n=1,2,3}
\zeta^n_R\zeta^n_L)\chi_R\chi_L\Big],
\label{Hmar12}
\end{equation}
where $g_1$ and $g_2$ are coupling constants.
Similar interactions appear for the (2-3) chains i.e., the $\pm B$ sector and 
also terms like $I^l_{AR}I^l_{BL}$, etc.
It is obvious that these interactions also generate mass matrix of the multiplyers 
$\lambda_{i\nu}$ through bubble diagrams.
We do not give detailed calculations of these diagrams but we believe that
in wide region of the parameter space of $(g_1, g_2)$ existence of the interactions 
(\ref{Hmar12}) etc. does not change the above result and excitations on the top
and bottom chains are described by the two massless scalar fields $\varphi_{\pm C}$.

Let us consider how the gapless modes on the top and bottom chains
evolve in the presence of inter-chain interactions between these chains,
i.e., we add the following term to the Hamiltonian,
\begin{equation}
H'_{rung}=J''\sum_i \vec{S}_{i,1}\vec{S}_{i,3}.
\label{H'rung}
\end{equation}
From the discussion given so far, one may think that the gapless modes acquire a
gap and low-energy excitations are described by spin-triplet Majorana
fermions.
However here again careful study on the zero-modes of the bosons
of the bosonization is required since the $\theta$-terms of the $A,B,C$-sectors
are not independent with each other.
It is easily proved that $\theta$'s must satisfy constraints similar 
to Eq.(\ref{const1}),
\begin{eqnarray}
\theta_{+A}-\theta_{-A} &=& \theta_{+B}-\theta_{-B},   \nonumber  \\
\theta_{+B}+\theta_{-B} &=& \theta_{+C}+\theta_{-C},  \label{thetaconst}  \\
\theta_{+C}-\theta_{-C} &=& \theta_{+A}+\theta_{-A}.  \nonumber 
\end{eqnarray}
From (\ref{thetaconst}), it is obvious that for $\theta_{\pm A}=\theta_{\pm B}=\pm
{\pi \over 2}$ we have $\theta_{\pm C}=0$ (mod $\pi$) and therefore the fermions
$\Psi_{\pm C}$ remain {\em massless} in the $\theta$-vacuum for $J''=0$ even if
the interaction (\ref{H'rung}) exists. 

It is expected that the $\theta$-vacuum for $J''=0$ tends to unstable 
because of the interaction (\ref{H'rung}) and
there exists some critical coupling $J''_c$ at which values of the parameter
$\theta$'s of the lowest energy change from those of the $J''=0$ case.
In that new phase, fermions $\Psi_{\pm C}$ acquire a mass.
The critical coupling $J''_c$ can be estimated from the low-energy
effective action of the 3-leg system similar to Eq.(\ref{Saux}) of the 2-leg case.
We expect that the charge sectors are less sensitive to the $J''$ coupling 
than the spin sectors since the charge-sector scalar fields are already massive
and their leading-order contributions to the effective potential of $\sigma$'s
are cancelled by the quadratic terms of $\sigma$'s similar to
the last term in (\ref{Saux}) $\sigma\sigma'$. 
Therefore let us study effective potential obtained from the spin sectors
and get a rough estimate of $J''_c$.

The effective potental of the auxiliary fields $\sigma'_X \; (X=A,B,C)$
or the masses $M_X$ are obtained by integrating over the fermions 
$\Psi_{\pm X}$.
To this end we ignore the constraints (\ref{const3}) as 
we can expect that  these constraints are irrelevant for low-energy excitations.
From (\ref{thetaconst}) independent variables are, e.g., $\theta_{+A},
\theta_{+B}$ and $\theta_{+C}$.
We futhermore require the symmetry between the first and the third chains,
i.e., we take $\theta_{+A}=\theta_{+B}$.
Then the effective potential is obtained as follows by intergating over the fermions 
$\Psi_{\pm A,B,C}$;
\begin{eqnarray}
V^{\mbox{\small{eff}}}/\Big( c\Lambda^2L/8\pi\Big) =
 6\Big({M_{AB}\over \Lambda c}\Big)^2\Big[\ln \Big({M_{AB} \over \Lambda c}\Big)^2-1\Big]
 +2\Big({3M_{AB}\over \Lambda c}\Big)^2
 \Big[\ln \Big({3M_{AB} \over \Lambda c}\Big)^2-1\Big]  && \nonumber   \\
   +\; 3\Big({M_C\over \Lambda c}\Big)^2\Big[\ln \Big({M_C \over \Lambda c}\Big)^2-1\Big]
 +\Big({3M_C\over \Lambda c}\Big)^2\Big[\ln \Big({3M_C \over \Lambda c}\Big)^2-1\Big], \;\;  &&
\label{effectV}
\end{eqnarray}
where 
\begin{eqnarray}
 M_{AB} &=& {M_s \over 2}\cdot\Big(\sin \theta_{+A}+ \sin (-\theta_{+B}+\theta_{+C})
\Big),  \nonumber  \\
 M_C/M_{AB} &=& \Big(J''/J'\Big)\cdot {\sin \theta_{+C} \over 
\sin \theta_{+A}+ \sin (-\theta_{+B}+\theta_{+C})}.
\label{massess}
\end{eqnarray}
From (\ref{effectV}), (\ref{massess}) and the fact that $f(x)=x(\ln x -1)$
is a decreasing function for $x<1$, 
it is obvious that $V^{\mbox{\small{eff}}}$
has the minimum at $\theta_{+A}=\theta_{+B}=\pm{\pi \over 2}$ and $\theta_{+C}=\pm\pi$
for $J''=0$ as in the 2-leg case.  
We shall study the stability of this $\theta$-vacuum in the presence of the 
$J''$ coupling.

We have numerically studied the effective potential $V^{\mbox{\small eff}}$ 
in the whole region of $\theta_{+A}$ and $\theta_{+C}$ and found that the absolute minimum
stays at $(\theta_{+A}, \theta_{+C})=({\pi \over 2}, \pi)$ for small values of $J''$.
However at some value of $J''$, the absolute minimum starts to 
move away from that point continuously. 
In order to study this behavior more precisely,
let us introduce variable $\delta$'s and parametrize $\theta$'s as,
\begin{equation}
\theta_{+A}=\pi/2 +\delta_{A}, \; \; \theta_{+B}=\pi/2 +\delta_{B},  \;\;
\theta_{+C}=\pi +\delta_{C}, \;\; \delta_{A,B,C} \ll 1.
\label{deltas}
\end{equation}
We assume $(A-B)$ symmetry, i.e., $\delta_A=\delta_B$. 
Then the masses are given as a function of $\delta$'s as 
\begin{eqnarray}
&& M_{AB} \propto \cos (\delta_A)+\cos (\delta_A-\delta_C) \sim 2-
\Big(\delta_A-{\delta_C \over 2}\Big)^2-{1 \over 4}\delta^2_C, \nonumber \\
&& M_C \propto -\sin (\delta_C).
\label{massAB}
\end{eqnarray}
From (\ref{massAB}), it is expected that the absolute minimum exists close to
the line $\delta_A=\delta_C/2$. 
Then let us study $V^{\mbox{\small eff}}$ on the line $\delta_A=\delta_C/2$.
Numerical calculations are shown in Fig.1.

The calculation of the effective potential in Fig.1 indicates
that transition to the phase with nonzero $\theta_{\pm C}$ (mod $\pi$) 
actually occures at a certain valu of $J''$.
Though this result is obtained by a rough estimate of the effective
potential as we have ignored
contribution from the charge sectors and the constraints (\ref{const3}),
we expect that there really exists some critical value of $J''$ at which phase transition
occures and fermions $\Psi_{\pm C}$ become massive.
It is possible that $J''_c=0$ or $J''_c=J'$ for some values of $J$.
If $J''_c<J'$, low-energy excitations are spin-triplet magnon and spin-singlet
excitation and ratio of their gaps is $3$.

Very recently symmetric antiferromagnetic case of the above system  
(i.e., $J'=J''$) was studied by Kawano and Takahashi
in Ref.\cite{takahashi} and they obtain the result that the system has an
excitation energy gap for any finite value of $J'=J''$.
This result supports the above prediction.

Let us take a brief look at the strong-inter-chain coupling to observe
the above transition in this limit.
We assume $J',J''>J$ and concentrate on the states on each rung.
There are three sites and totally eight states on each rung. 
Eigen states and eigen values are given as follows;
\begin{eqnarray}
&& |\alpha\rangle_1|S\rangle_{23}-|S\rangle_{12}|\alpha\rangle_3, 
 \;\; E=-J'+{J'' \over 4}, \nonumber  \\
&& |S\rangle_{13}|\alpha\rangle_2,  
\;\; E=-{3J'' \over 4}, \;\;(\alpha=\uparrow, \downarrow) \nonumber  \\
&& |\uparrow\uparrow\downarrow\rangle_{123} +|\uparrow\downarrow\uparrow\rangle_{123}
+|\downarrow\uparrow\uparrow\rangle_{123}, \; \; (\uparrow \leftrightarrow \downarrow),
\;\; E={J' \over 2}+{J'' \over 4}, \label{rungstates}  \\
&& |\uparrow\uparrow\uparrow\rangle_{123}, \; \; 
|\downarrow\downarrow\downarrow\rangle_{123},  
\;\; E={J' \over 2}+{J'' \over 4}, \nonumber
\end{eqnarray}
where $|S\rangle_{ij}\equiv |\uparrow\rangle_i|\downarrow\rangle_j-
|\downarrow\rangle_i|\uparrow\rangle_j$ and the other notations are self-evident.
Therefore for $J'>J''$, the lowest-energy states are doubly degenerate
and they have spin-${1 \over 2}$ degrees of freedom.
At finite $J$ coupling, this spin-${1 \over 2}$ degrees of freedom on each rung
behave like spin on a single spin chain.
This is the reason why the system has no energy gap (in the 
strong-inter-chain couplig limit) and behaves like the spin-${1 \over 2}$ chain.
On the other hand at $J'=J''$, degeneracy of the lowest-energy states doubles
because of the permutation symmetry.
Therefore at finite $J$, one can expect a phase transition near this point.
Actually the system is maximally frustrated at $J'=J''$ and 
the lowest-energy state has rather complex structure\cite{takahashi}. 
Our analysis in this paper suggests that this transtion remains in the 
weak-inter-chain coupling region.

\section{Conclusion}
\setcounter{equation}{0}
In this paper, we discussed low-energy excitations in the AF Heisenberg model on
2-leg and 3-leg ladders.
We used gauge-theoretical description of quantum spin chain and bosonization
method.
It was shown that in the 2-leg system low-energy excitations are described by
spin-triplet and spin-singlet  Majorana fermions.
In order to obtain this result, the zero-modes in the bosonization play an
important role.
Our study justifies the results obtained by Shelton et.al. in Ref.\cite{shelton}.
Then we apply the same method to the 3-leg case and showed that  excitations
on the top and bottom chains are described by two massless scalar fields.

One may wonder how our method is applied to the 4-leg system.
In order to obtain a low-energy effective field theory, we first integrate out
high-momentum modes in the original lattice system.
In the 2-leg and 3-leg cases, this procedure is rather straightforward.
However in the 4-leg case, it is naturally expected that nontrivial
renormalization of the inter-chain couplings appears, i.e., 
the couplings $J_{12}$ and $J_{34}$ develop and $J_{23}$ tends
to small.
In low-energy states, spin-singlet pairs form on the $(1-2)$ and  $(3-4)$
ladders respectively.
Then it is natural to describe low-energy excitations on the 4-leg ladder
by starting with the system of two decoupled 2-leg ladders.
At present this problem is under study and results will be reported in a future publication.

\newpage
\appendix
\renewcommand{\theequation}{\Alph{section}.\arabic{equation}}
{\Large {\bf Appendix}}
\setcounter{equation}{0}
\section{Explicit form of $C$'s}
In this appendix, we shall give one of the possible choice of operators $C$
which appear the bosonization formulas (\ref{bosonizedF1}) and (\ref{bosonizedF2}),
\begin{eqnarray}
  C_{1 R \uparrow}
    &=& \exp \bigl[ \pi i\bigl( p_{1 L \uparrow}
       + p_{1 L \downarrow} + p_{2 \uparrow}
       + p_{2 L \downarrow} \bigr)
       +  4\pi  ip_{1 L \uparrow} \bigr]
        \nonumber \\
    &=& \exp \bigl[2\pi  ip_{- C L}
        + 2\pi  i\bigl( p_{+ C L} + p_{- C L}
        + p_{+ S L} + p _{- S L} \bigr) \bigr],
        \nonumber \\
  C_{1 R \downarrow}
    &=& C_{2 R \uparrow} \nonumber \\
    &=& \exp \bigl[\pi i\bigl( p_{1 L \uparrow}
       + p_{1 L \downarrow} + p_{2 \uparrow}
       + p_{2 L \downarrow} \bigr) 
    +  \pi i\bigl( p_{1 R \downarrow}
       + p_{2 R \uparrow} \bigr) \bigr]
       \label{eq:CR} \\
    &=& \exp \bigl[ 2\pi  ip_{- C L}
        +  \pi i\bigl( p_{+ C R} + p_{+ S R}
       \bigr) \bigr], \nonumber \\
  C_{2 R \downarrow}
    &=& \exp \bigl[ \pi i\bigl( p_{1 L \uparrow}
       + p_{1 L \downarrow} + p_{2 \uparrow}
       + p_{2 L \downarrow} \bigr)  
    +  \pi i\bigl( p_{1 L \uparrow}
       + p_{1 L \downarrow} - p_{2 L \uparrow}
        - p_{2 L \downarrow} \bigr) \bigr]
        \nonumber \\
    &=& \exp \bigl[ 2\pi  ip_{- C L}
        +  2\pi  ip_{- C R} \bigr],
        \nonumber
\end{eqnarray}
\begin{eqnarray}
  C_{1 L \uparrow} &=& 1, \nonumber \\
  C_{1 L \downarrow}
    &=& C_{2 L \uparrow} \nonumber \\
    &=& \exp \bigl[  \pi i\bigl( p_{1 L \downarrow}
       + p_{2 L \uparrow} \bigr) \bigr] \nonumber \\
    &=& \exp \bigl[  \pi i\bigl( p_{+ C L} + p_{+ S L}
       \bigr) \bigr],
       \label{eq:CL} \\
  C_{2 L \downarrow}
    &=& \exp \bigl[  \pi i\bigl( p_{1 L \uparrow}
       + p_{1 L \downarrow} - p_{2 L \uparrow}
        - p_{2 L \downarrow} \bigr) \bigr]
        \nonumber \\
    &=& \exp \bigl[  2\pi i p_{- C L} \bigr].
        \nonumber
\end{eqnarray}
Then for spin and charge fermions,
\begin{eqnarray}
    C_{+ C R}
    &=& \exp \bigl[\pi i\bigl(p_{+CL} + p_{-CL}
       + p_{+SL} + p_{-SL} \bigr) + i 4\pi  p_{- C L}
       \nonumber \\
    & & {} + \pi i\bigl( p_{+ C R} + p_{- C R}
       + p_{+ S R} \bigr) \bigr],
       \nonumber \\
  C_{- C R}
    &=& \exp \bigl[ \pi i\bigl(p_{+CL} + p_{-CL}
       + p_{+SL} + p_{-SL} \bigr)
           -  \pi ip_{- C R} \bigr],
        \nonumber \\
  C_{+ S R}
    &=& \exp \bigl[ \pi i\bigl(p_{+CL} + p_{-CL}
       + p_{+SL} + p_{-SL} \bigr)
          -  \pi ip_{- C R} \bigr],
        \label{eq:nCR} \\
  C_{- S R}
    &=& \exp \bigl[ \pi i\bigl(p_{+CL} + p_{-CL}
       + p_{+SL} + p_{-SL} \bigr)
       \nonumber \\
    & & {} +  \pi i\bigl(- p_{+ C R}
       + p_{- C R} - p_{+ S R} \bigr) \bigr],
       \nonumber
\end{eqnarray}

\begin{eqnarray}
  C_{+ C L}
    &=& \exp \bigl[ \pi i\bigl( p_{+ C L}
       + p_{- C L} + p_{+ S L} \bigr) \bigr],
       \nonumber \\
  C_{- C L}
    &=& \exp \bigl[-  \pi ip_{- C L} \bigr],
        \nonumber \\
  C_{+ S L}
    &=& \exp \bigl[-  \pi ip_{- C L} \bigr],
        \label{eq:nCL} \\
  C_{- S L}
    &=& \exp \bigl[ \pi i\bigl(- p_{+ C L}
       + p_{- C L} - p_{+ S L} \bigr) \bigr].
       \nonumber
\end{eqnarray}

\newpage

{\Large Figure Caption}  \\

The effective potentials $V^{\mbox{eff}}$ as a function of $\delta_A$,
for  $J''/J'=0.3, 0.6$ and $0.9$.
It is observed that the minimum exists at $\delta_A=0$ at small values of $J''/J'$,
whereas at some value of $J''/J'$ the point $\delta_A=0$ becomes unstable.
Phase transition occures and the system has an energy gap in the new phase.


\newpage


\begin{thebibliography}{1}

\bibitem{rice}For review, see E.Dagotto and T.M.Rice,
Science271(1996)618.

\bibitem{NLS}D.Senechal,
Phy.Rev.B52,15319(1995);  \\
G.Sierra, J.Phys.A29,3299(1996);  \\
S.Dell'Aringa, E.Ercolessi, G.Morandi, P.Pieri and M.Roncaglia,  \\
Phys.Rev.Lett.78,2457(1997);  \\
I.Ichinose and T.Matsui, preprint UT-Komaba 97-13 cond-mat/9710002,
{\it ``Quasi-excitations and superconductivity in the t-J model on a ladder"}.

\bibitem{shelton}D.G.Shelton, A.A.Nersesyan and A.M.Tsvelik,
Phys.Rev.B53(1996)8521.

\bibitem{kishine}J.Kishine and H.Fukuyama,
J.Phys.Soc.Japan 66(1997)26.

\bibitem{hubbard}L.Balents and M.P.A.Fisher,
Phys.Rev.B53(1996)12133;  \\
E.Arrigoni, Phys.Lett.A215(1996)91;  \\
H-H. Lin, L.Balents and M.P.A.Fisher; preprint cond-mat/9703055,
{\it ``The N-chain Hubbard model in weak coupling"}.

\bibitem{hosotani}Y.Hosotani,
preprint cond-mat/9707129, {\it ``Gauge theory description of spin ladders"}.

\bibitem{mukaida}C.Itoi and H.Mukaida,
J.Phys.A27(1994)4695.

\bibitem{boson} See for example, E.Abdalla, M.C.B.Abdalla and K.D.Rothe,
{\it ``2 Dimensional Quantum Field Theory"} (World Scientific 1991).

\bibitem{RGSG}I.Ichinose and H.Mukaida,
Int.J.Mod.Phys.A9(1994)1043.

\bibitem{HH}J.E.Hetrick and Y.Hosotani,
Phys.Rev.D38(1988)2621.

\bibitem{takahashi}K.Kawano and M.Takahashi,
preprint cond-mat/9709271, 
{\it ``Three-leg antiferromagnetic Heisenberg ladder with frustrated boundary
condition; Ground state properties"}.


\end{thebibliography}
\end{document}